# The Greek Public Debt Path

*From Zero to Infinity*

**Dimitris Sardelis**

Athens, Greece, December 2, 2012

### An Overview

The aim of the present article is to treat the Greek public debt issue strictly as a curve fitting problem. Thus, based on Eurostat data and using the *Mathematica* technical computing software, an exponential function that best fits the data is determined modelling how the Greek public debt expands with time. Exploring the main features of this best fit model, it is concluded that the Greek public debt cannot possibly be serviced in the long run unless a radical growth is implemented and/or part of the debt is written off.

## Basic Facts

### Prerequisite Remarks

- The subject of the present article is the Greek Public Debt. Having tried hard to understand what some economists have to say on the subject, their utmost certainty and admirable confort did not convince me at all because I could hardly disentangle the mere facts from philosophical pre-conceptions and political projections. Consequently, I decided to confine myself to an understanding of the issue from a strictly mathematical, statistical point of view.
- The present article is based on Eurostat data for Greece for the period 1995 to 2011 (Tables 1A and 1B below), during which the figures given are beyond dispute.
- For the editing and the Statistical Model Analysis also involving predictions, exclusive use is made of the *Mathematica* technical computing software. Every *Mathematica* built-in function that is employed, is also displayed in the exposition so that the derived results are reproducible and, consequently, they can be verified or be falsified by almost anyone.

### Abbreviations

**t:** years after 1995
**PD**: Public Debt (in Eurostat terms: General government gross debt) in billion euros
**GDP**: Annual Gross Domestic Product in billion euros
**PR**: Annual Public Revenue (in Eurostat terms: Total general government revenue) in billion euros
**PE**: Annual Public Expenditure (in Eurostat terms: Total general government expenditure) in billion euros
**PN**: **PR−PE** Annual Net Public Profit [Surplus (+) or Deficit (−)] (in Eurostat terms: General government deficit/surplus) in billion euros

### Eurostat Data

Eurostat[1] provides separate tables for the percentage ratios PD/GDP (tsd410), PR/GDP (tec00021), PE/GDP (tec00023), PN/GDP (tec00127) as well as for PD (tsd410) of all (27) European Union countries.

The percentage figures for Greece are displayed in Table 1A below:

**Table 1A**

| Year | t | PD/GDP (%) | PR/GDP (%) | PE/GDP (%) | PN/GDP (%) |
|------|----|-----------|-----------|-----------|-----------|
| 1995 | 0  | 97.   | 36.7 | 45.7 | -9.   |
| 1996 | 1  | 99.4  | 37.4 | 44.1 | -6.7  |
| 1997 | 2  | 96.6  | 39.  | 44.9 | -5.9  |
| 1998 | 3  | 94.5  | 40.5 | 44.3 | -3.8  |
| 1999 | 4  | 94.   | 41.4 | 44.5 | -3.1  |
| 2000 | 5  | 103.4 | 43.  | 46.7 | -3.7  |
| 2001 | 6  | 103.7 | 40.9 | 45.4 | -4.5  |
| 2002 | 7  | 101.7 | 40.3 | 45.1 | -4.8  |
| 2003 | 8  | 97.4  | 39.  | 44.7 | -5.7  |
| 2004 | 9  | 98.6  | 38.1 | 45.5 | -7.4  |
| 2005 | 10 | 100.  | 39.  | 44.6 | -5.6  |
| 2006 | 11 | 106.1 | 39.2 | 45.3 | -6.1  |
| 2007 | 12 | 107.4 | 40.7 | 47.5 | -6.8  |
| 2008 | 13 | 112.9 | 40.7 | 50.6 | -9.9  |
| 2009 | 14 | 129.7 | 38.3 | 54.  | -15.7 |
| 2010 | 15 | 148.3 | 40.6 | 51.5 | -10.9 |
| 2011 | 16 | 170.6 | 42.3 | 51.8 | -9.5  |

The GDP, PR, PE, and PN figures for Greece listed in Table 1B below, are deduced from the aforementioned Eurostat tables.

**Table 1B**

| Year | t | PD | GDP | PR | PE | PN |
|------|----|---------|---------|---------|---------|----------|
| 1995 | 0  | 95.0133 | 97.9519 | 35.9483 | 44.764  | -8.81567 |
| 1996 | 1  | 107.666 | 108.316 | 40.5103 | 47.7675 | -7.25719 |
| 1997 | 2  | 114.864 | 118.907 | 46.3738 | 53.3893 | -7.01552 |
| 1998 | 3  | 115.679 | 122.412 | 49.5767 | 54.2284 | -4.65164 |
| 1999 | 4  | 122.335 | 130.143 | 53.8793 | 57.9138 | -4.03444 |
| 2000 | 5  | 140.971 | 136.336 | 58.6243 | 63.6687 | -5.04442 |
| 2001 | 6  | 151.869 | 146.45  | 59.8982 | 66.4885 | -6.59027 |
| 2002 | 7  | 159.214 | 156.553 | 63.0907 | 70.6052 | -7.51453 |
| 2003 | 8  | 168.025 | 172.51  | 67.279  | 77.1121 | -9.83309 |
| 2004 | 9  | 183.157 | 185.758 | 70.7736 | 84.5197 | -13.7461 |
| 2005 | 10 | 195.421 | 195.421 | 76.2142 | 87.1578 | -10.9436 |
| 2006 | 11 | 224.204 | 211.314 | 82.835  | 95.7252 | -12.8901 |
| 2007 | 12 | 239.3   | 222.812 | 90.6845 | 105.836 | -15.1512 |
| 2008 | 13 | 263.284 | 233.201 | 94.9128 | 118.    | -23.0869 |
| 2009 | 14 | 299.682 | 231.058 | 88.4951 | 124.771 | -36.2761 |
| 2010 | 15 | 329.513 | 222.194 | 90.2106 | 114.43  | -24.2191 |
| 2011 | 16 | 355.658 | 208.475 | 88.1848 | 107.99  | -19.8051 |

# The Best Fit Model for the PD Path

The function that best fits the data for the PD over the years, can be determined by employing any of the mathematics-statistic software packages currently in use. The PD data trend suggests that a strong function candidate to be explored is the exponential function of the form $pe^{rt}$, where *p* and *r* are data dependent constants.

In all that follows, we shall use extensively the *Mathematica* technical computing software [**2**].

## Statistical Model Analysis

Applying the *Mathematica* built-in functions

```
nlmgreekpd = NonlinearModelFit[greekpd, p Exp[r t], {p, r}, t];
{nlmgreekpd["ParameterConfidenceIntervalTable"], nlmgreekpd["ANOVATable"]}
```

one gets the following summary report:

| | Estimate | Standard Error | Confidence Interval |
|---|---|---|---|
| p | 88.375   | 2.4402    | {83.1738, 93.5761}   |
| r | 0.085893 | 0.00224162 | {0.0811151, 0.0906709} |

| | DF | SS | MS |
|---|----|---------|---------|
| Model            | 2  | 731243. | 365621. |
| Error            | 15 | 903.653 | 60.2435 |
| Uncorrected Total | 17 | 732146. | |
| Corrected Total  | 16 | 104745. | |

**Model Accuracy**

The degree of accuracy (scale 0 to 100 percent) for any statistical non linear model, may be expressed by the coefficient of determination $R^2$ defined as the ratio of the difference between the uncorrected total sum of squares and the residual sum of squares to the uncorrected total sum of squares, both quantities being displayed in the ANOVA table for the tested functional model. The coefficient of determination for the best fit exponential model is

$$R^2 = (732\,146 - 903.653)/732\,146 = 0.998766 \text{ or } 99.88\,\%$$

(1) can also be derived directly by a built-in *Mathematica* function:

```
nlmgreekpd["RSquared"]
0.998766
```

To illustrate the measure of absolute excellence (100 percent) for statistically founded exponential models, it is worth noting that if PD data consisted of two points only, e.g., any two points (0, *a*) and (*n*, *b*), the coefficient of determination $R^2$ for the exponential curve passing through both points would be 100 percent.

## Function of Best Fit

From the summary report displayed above, the function of best fit is

$$P(t) = 88.375\, e^{0.085893\, t} \quad (1)$$

*P(t)*: Expected Public Debt (billion euros), and *t*: years after 1995.

Differentiating (1) with respect to time, the percentage rate of *P(t)* is found to be a constant:

$$\frac{1}{P}\frac{dP}{dt} = 0.085893 \text{ or } 8.59\,\% \quad (2)$$

Thus, it may be said that the expected PD expands annually at the constant rate of 8.59 percent.

Taking into account that the typical growth rates of countries (see [**3**]) rarely exceed the expansion rate (2), it looks more than certain that the actual Greek PD can never be serviced, exactly as a latecomer passenger can never catch-up his plane just before takeoff.

The expansion rate (**2**) can be used to determine how long it takes for the expected PD to reach multiple values. Let

$$P(t + T_n) = n\, P(t)$$

for any *n* and *t*. Then (1) yields

$$T_n = \frac{Ln\,[n]}{0.085893} \quad (3)$$

This formula serves to convert the annual expansion rate of 8.59 percent into time units. Applying (3), it is straightforward to find that it takes 8.1 years for the expected PD to double, 12.8 years to triple, e.t.c.

## Actual and Expected PD

The exponential function (1) can also be used to predict the PD for different years. Applying the *Mathematica* built-in function

```
nlmgreekpd["SinglePredictionConfidenceIntervalTable", ConfidenceLevel → .95]
```

we deduce the Table 2 below for the years 1995 to 2020:

**Table 2**

```
         Year    t    ObservedPD    PredictedPD    SE         CILow      CIHigh
         1995    0    95.0133       88.375         8.13622    71.033     105.717
         1996    1    107.666       96.3013        8.14104    78.9491    113.654
         1997    2    114.864       104.939        8.14161    87.5851    122.292
         1998    3    115.679       114.35         8.13745    97.0059    131.695
         1999    4    122.335       124.606        8.12826    107.282    141.931
         2000    5    140.971       135.782        8.11411    118.488    153.077
         2001    6    151.869       147.961        8.09556    130.705    165.216
         2002    7    159.214       161.231        8.07399    144.022    178.441
         2003    8    168.025       175.692        8.05196    158.53     192.854
         2004    9    183.157       191.45         8.0337     174.326    208.573
         2005    10   195.421       208.621        8.02574    191.514    225.727
         2006    11   224.204       227.332        8.03774    210.2      244.464
         2007    12   239.3         247.721        8.08336    230.492    264.951
         2008    13   263.284       269.939        8.18112    252.502    287.377
         2009    14   299.682       294.15         8.35507    276.342    311.959
         2010    15   329.513       320.532        8.63479    302.128    338.937
         2011    16   355.658       349.281        9.05448    329.982    368.58
         2012    17   -             380.608        9.65108    360.037    401.179
         2013    18   -             414.744        10.4619    392.445    437.043
         2014    19   -             451.943        11.5225    427.383    476.502
         2015    20   -             492.477        12.8661    465.054    519.901
         2016    21   -             536.647        14.5235    505.691    567.604
         2017    22   -             584.779        16.5253    549.556    620.002
         2018    23   -             637.228        18.9029    596.937    677.519
         2019    24   -             694.381        21.6911    648.147    740.614
         2020    25   -             756.659        24.9284    703.526    809.793
```

Table 2 displays the observed PD, the predicted/expected PD and the corresponding ninety five percent confidence PD-intervals, i.e., the range of the expected predictions for each year. The uncertainty of the predictions increases with time and it is reflected in the expanding confidence interval limits.

Let us outline the essential in these predictions: Provided that the exponential model trend continues beyond the current domain, the expected PD will exceed on average 400 billion euros by the year 2013, 450 billion euros by the year 2014 and 750 billion euros by the year 2020!

Figure 1 graphically displays the function (1), the PD data at disposal and the bands set by a ninety five percent degree of confidence:

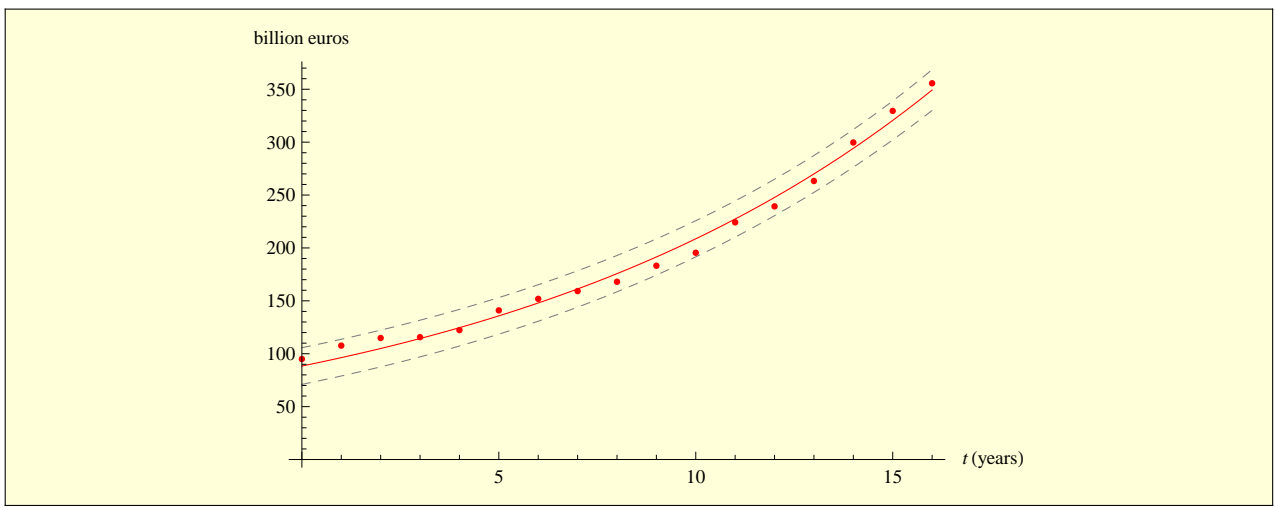

Figure1: PD Data and P(t) with 95 percent Confidence Bands

### An Optimistic Estimate for the PD to GDP ratio

To further illustrate the predictions scheme, let us consider the predicted ninety five per cent confidence interval of the PD for the year 2014 (see Table 2). This interval can be expressed in the form **451.943±24.5595** billion euros. Note that the highest GDP ever recorded took place in the year 2008 (see Table 1). Since then, the GDP has been declining steadily. Subsequently, even if the GDP by the year 2014 reverses its course and returns to its 2008 highest value, the best possible, most optimistic confidence interval for the PD to GDP ratio will then be **193.8 ±10.53** percent, i.e., the PD to GDP ratio will still be far above its 170.6 percent 2011 value.

# PD Sustainability

On November 26, 2012 the Eurozone Finance Ministers and the Internation Monetary Fund (IMF) have set a 124 percent threshold for the Greek PD to GDP ratio by the year 2020 so that PD can be considered sustainable [**4**].

How possible is that to happen? More specifically, in order for this threshold to be reached by 2020, (A) which GDP growth rate is it required if the PD expansion trend remains unaltered? (B) how much of PD should be written off if the GDP growth trend remains unaltered? (C) If both the PD and GDP trends remain unaltered, what is the probability that the PD to GDP ratio will then be at most 124 per cent?

## (A) By Re-Starting Growth

Let the Greek GDP grow exponentially according to the law $f(t) = a\, e^{b\,t}$ so that

$$\frac{P(16)}{f(16)} = 1.706 \text{ and } \frac{P(25)}{f(25)} = 1.24 \tag{4}$$

Then, by solving the system of equations (4), one finds that $a = 29.3783$ and $b = 0.121342$.

The graphs of the functions $P(t)$ and $f(t)$ are displayed in Figure 2 below:

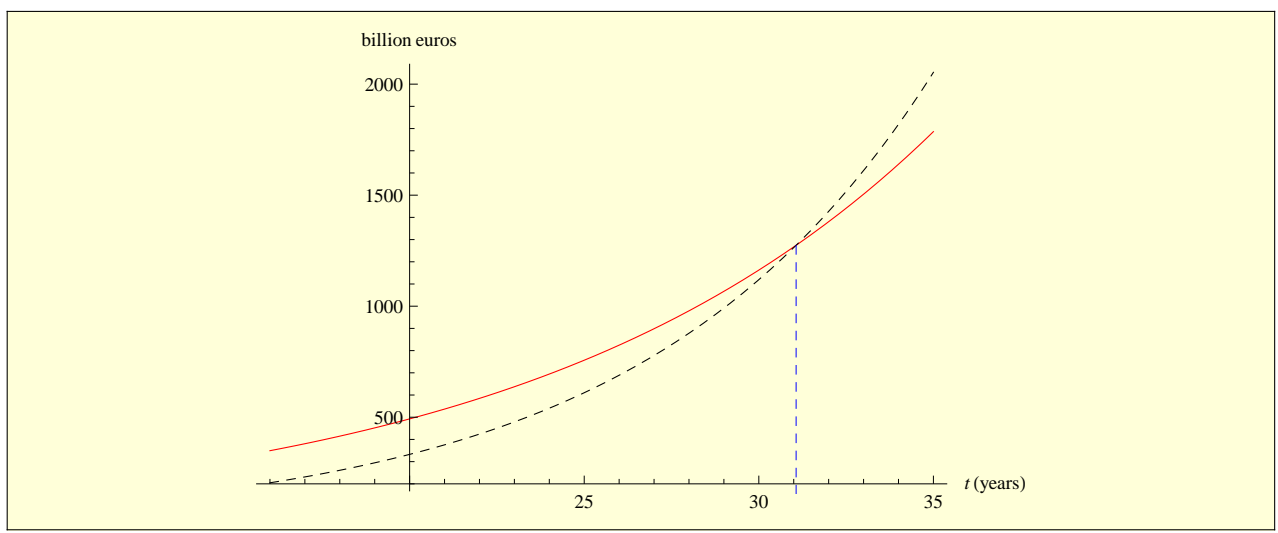

Figure2: P(t) and f(t)

Therefore, if the 124 percent threshold for the PD to GDP ratio is to be reached by the year 2020, the GDP must attain a growth rate of 12.1 percent. The feasibility of this seems very unlikely mainly because the 2011 percentage GDP change was

$$\frac{GDP[2011] - GDP[2010]}{GDP[2010]} = -0.0617433 \text{ or } -6.2\,\%$$

and the resulting growth rate gap of 18.3 percent cannot be easily bridged.

Nevertheless, if everything goes according to schedule, a direct consequence of this scenario is that the PD to GDP ratio would drop below 100 per cent after the year 2027, i.e., beyond the intersection point of the $P(t)$ and $f(t)$ curves shown in Figure 2. In other words, since $P(t) < f(t)$ for all $t > 32$, Greece after 2027 would return to a steady surplus period.

## (B) By Writing Off Part of PD

This appears to be a positive and , as we shall see, a quite feasible scenario for the temporary servicing of the Greek Public Debt.

Let the expected PD and GDP grow exponentially according to the laws $g(t) = c\, e^{d\,t}$ and $Q(t) = q\, e^{k\,t}$ respectively, so that

$$\frac{g(18)}{P(18)} = 1 - x, \quad \frac{g(18)}{Q(18)} = y, \quad \frac{g(25)}{Q(25)} = 1.24, \tag{5}$$

where $x$ denotes the PD proportion to be written off in the year 2013 ($0 < x < 1$), and $y$ denotes the corresponding PD to GDP ratio ($y > 0$).

To determine $Q(t)$ that best fits the GDP data (see Table 1B), we can apply as before the *Mathematica* built-in functions

```
nlmgreekgdp = NonlinearModelFit[greekgdp, q Exp[k t], {q, k}, t];
```

```
{nlmgreekgdp["ParameterConfidenceIntervalTable"], nlmgreekgdp["ANOVATable"],
 {RSquared, nlmgreekgdp["RSquared"]}}
```

and get

|   | Estimate | Standard Error | Confidence Interval |
|---|---|---|---|
| q | 111.232 | 5.8758 | {98.7077, 123.756} |
| k | 0.0499389 | 0.00469206 | {0.039938, 0.0599398} |

,

|   | DF | SS | MS |
|---|---|---|---|
| Model | 2 | 525661. | 262831. |
| Error | 15 | 3626.12 | 241.742 |
| Uncorrected Total | 17 | 529287. |   |
| Corrected Total | 16 | 34646.2 |   |

, {RSquared, 0.993149}

Thus, the expected GDP is found to be

$$Q(t) = 111.232 \, e^{0.0499389 \, t} \tag{6}$$

After inserting (6) into (5), the system of equations can be solved for *x*, *c* and *d* in terms of *y*.

Table 3 below displays the solutions for some reasonable *y*−values that are on track with the 124 percent threshold.

**Table 3**

| y(%) | x | c | d | g(18) |
|---|---|---|---|---|
| 150. | 0.0116118 | 272.207 | 0.0227455 | 409.929 |
| 145. | 0.044558 | 241.166 | 0.0275886 | 396.264 |
| 140. | 0.0775043 | 212.759 | 0.0326016 | 382.6 |
| 135. | 0.110451 | 186.844 | 0.037797 | 368.936 |
| 130. | 0.143397 | 163.284 | 0.0431885 | 355.272 |
| 124. | 0.182932 | 137.928 | 0.0499389 | 338.874 |
| 120. | 0.209289 | 122.685 | 0.0546232 | 327.943 |
| 115. | 0.242236 | 105.385 | 0.0607031 | 314.279 |
| 110. | 0.275182 | 89.9153 | 0.0670534 | 300.614 |
| 105. | 0.308128 | 76.1515 | 0.0736991 | 286.95 |
| 100. | 0.341075 | 63.9739 | 0.0806691 | 273.286 |

Thus, within the feasible region between 100 and 150 percent of the PD to GDP ratio (*y%*), PD in the year 2013 should be reduced by 1 to 34 percent (*100 x*) so that the 124 percent threshold is attained in the year 2020.

Let it be observed from Table 3 that *g(18)* ≥ *Q(18)*, in agreement with (5). Then, there are two possible cases: (i) For *y* ≥ *1.24*, the *g(t)* rates are less than or equal to to the rate of *Q(t)*. Subsequently, their corresponding curves must get closer to each other with time. (ii) For *y* < *1.24*, the *g(t)* rates are greater than the rate of *Q(t)* and the curves of *g(t)* and *Q(t)* must get further apart with time.

Figure 3 displays the graphs of *Q(t)* and of two forms for *g(t)*, one for each case (i) and (ii). The *y%* for case (i) is 140 percent and for case (ii) 110 percent.

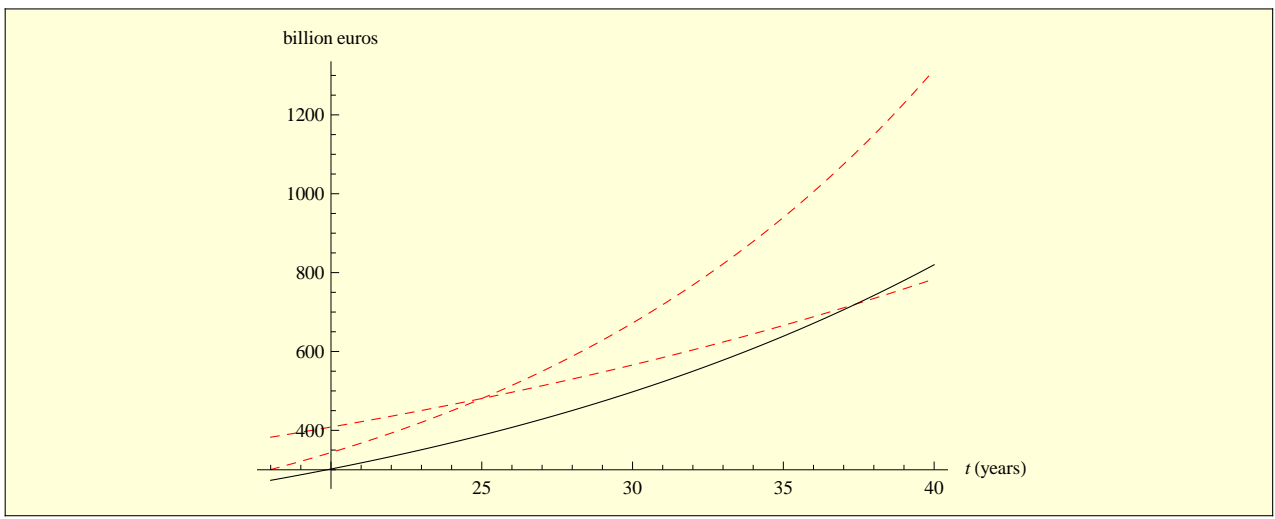

Figure3: Q(t) and g(t) [cases (i) and (ii)]

Note that the graph of *g(t)* for case (i) intersects the graph of *Q(t)* at a point beyond which *Q(t)* ≻ *g(t)*. It can be found easily that this turnover point for the g-function displayed, is expected to take place in the year 2033.

## (C) By Changing Neither

If both the PD and GDP trends remain unaltered, so is that of the PD to GDP ratio. Let us determine the exponential function *R(t)* that best fits the PD to GDP data (see Table 1A). Applying the *Mathematica* built-in functions

```
nlmgreekpdtogdp = NonlinearModelFit[greekpdtogdp, R Exp[λ t], {R, λ}, t];
{nlmgreekpdtogdp["ParameterConfidenceIntervalTable"], nlmgreekpdtogdp["ANOVATable"],
 {RSquared, nlmgreekpdtogdp["RSquared"]}}
```

we find

|   | Estimate | Standard Error | Confidence Interval |
|---|---|---|---|
| R | 84.129 | 5.5021 | {72.4016, 95.8564} |
| λ | 0.031337 | 0.00616621 | {0.018194, 0.0444799} |

|   | DF | SS | MS |
|---|---|---|---|
| Model | 2 | 208148. | 104074. |
| Error | 15 | 2693.88 | 179.592 |
| Uncorrected Total | 17 | 210842. |   |
| Corrected Total | 16 | 7051.3 |   |

, {RSquared, 0.987223}

Thus, the function that best fits the data of the PD to GDP ratio has the form

$$R[t] = 84.129 \, e^{0.031337 \, t} \qquad (7)$$

Let us also derive the ninety five percent confidence for a single prediction at any time *t* by employing the *Mathematica* built-in function

```
cisingleprediction[t_] = {greekpdtogdpfit[t_], greekpdtogdpsinglep[t_]} =
   nlmgreekpdtogdp[{"BestFit", "SinglePredictionBands"}]
```

For the year 2020, we find

| Year | t | R (t) | SE (t) | CILow(t) | CIHigh(t) |
|---|---|---|---|---|---|
| 2020 | 25 | 184.155 | 22.7944 | 135.569 | 232.74 |

The third entry is the expected PD to GDP ratio, the next entry is the standard error, and the last two entries are the ninety five percent confidence limits.

The probability that any a particular ratio lies beyond a pre-assigned value can be derived by applying directly the *Mathematica* built-in function

```
Needs["HypothesisTesting`"]; StudentTPValue[ratio-R(t)/SE(t), d.f = Length[data] - 2]
```

For example, the probability that the PD to GDP ratio will be less that 135.69, is found to be 0.025. The probability that the PD to GDP ratio will be more than 232.74, is also found to be 0.025. Hence, the probability that the PD to GDP ratio will lie anywhere between 135.69 and 232.74 is 0.95, thus verifying that the prediction confidence interval has indeed been set to ninety five percent.

We are now ready to answer the (C) question. Starting from the single prediction summary report for *t* = 25 and using directly the *Mathematica* built-in function right above, we have that if both the PD and GDP trends remain unaltered, then the probability that the PD to GDP ratio will be at most 124 in the year 2020, is found to be 0.00971359.

# Summary

The aim of the present article was to offer some understanding of the Greek Public Debt from a strictly mathematical, statistical point of view. Based on Eurostat data (Tables 1A and 1B) and using exclusively the *Mathematica* technical computing software for statistical analysis and predictions, we determined the best fit exponential functions which reflect quite accurately the PD data trend [Equation (1)], the GDP data trend [Equation (6)] and the PD to GDP ratio data trend [Equation (7)].

Regarding the PD model, we derived a single prediction confidence interval table (Table 2) for the expected PD up to the year 2020 which was used for various estimates.

The PD sustainability issue was examined from three aspects: (A) by re-starting growth given that the PD trend remains unaltered, (B) by writing off part of PD given that the GDP trend remains unaltered, and (C) by finding the probability that the 124 percent threshold is reached given that both the PD and the GDP trends remain unalterd. Among the three scenarios, scenario (B) seems to be the most feasible.